\documentclass[%
 aip,
 amsmath,amssymb,
 preprint,%
]{revtex4-1}

\usepackage{graphicx}
\usepackage{dcolumn}
\usepackage{bm}

\usepackage[utf8]{inputenc}
\usepackage[T1]{fontenc}
\usepackage{mathptmx}
\usepackage{etoolbox}
\usepackage{multirow}

\makeatletter
\def\@email#1#2{%
 \endgroup
 \patchcmd{\titleblock@produce}
  {\frontmatter@RRAPformat}
  {\frontmatter@RRAPformat{\produce@RRAP{*#1\href{mailto:#2}{#2}}}\frontmatter@RRAPformat}
  {}{}
}%
\makeatother
\begin{document}

\preprint{AIP/123-QED}

\title{Study of Surface Damage in Silicon by Irradiation with Focused Rubidium Ions}
\author{S. Xu} 
\author{Y. Li}
\author{M. A. Verheijen}
    \altaffiliation[Also at ]{Eurofins Materials Science, High Tech Campus 11, 5656AE Eindhoven, The Netherlands}
\affiliation{ Department of Applied Physics, Eindhoven University of Technology, P.O. Box 513, 5600 MB Eindhoven, the Netherlands}%
\author{E. R. Kieft}
\affiliation{ Thermo Fisher Scientific, Achtseweg Noord 5, 5651 GG Eindhoven, the Netherlands}%
\author{E. J. D. Vredenbregt*}
\affiliation{ Department of Applied Physics, Eindhoven University of Technology, P.O. Box 513, 5600 MB Eindhoven, the Netherlands}%
    \email{e.j.d.vredenbregt@tue.nl}

\date{\today}

\begin{abstract}
Cold atom ion sources have been developed and commercialized as alternative sources for focused ion beams (FIB). So far, applications and related research have not been widely reported. In this paper, a prototype rubidium FIB is used to study the irradiation damage of 8.5 keV beam energy Rb$^+$ ions on silicon to examine the suitability of rubidium for nanomachining applications. Transmission electron microscopy combined with energy dispersive X-ray spectroscopy is applied to silicon samples irradiated by different doses of rubidium ions. The experimental results show a duplex damage layer consisting of an outer layer of oxidation without Rb and an inner layer containing Rb mostly at the interface to the underlying Si substrate. The steady-state damage layer is measured to be $23.2(\pm 0.3)$ nm thick with a rubidium staining level of $7(\pm1)$ atomic percentage.
\end{abstract}

\maketitle

\section{\label{sec:intro}Introduction}
Focused Ion Beam (FIB) techniques are widely applied in various research fields for imaging, nanomachining, and surface modification\cite{bassim2014recent,munroe2009application,li2021recent}. Ion milling is one of the common uses in the semiconductor industry, such as for specific layer removal, cross-sectioning, and transmission electron microscopy (TEM) sample preparation\cite{li2006recent}. However, during the FIB treatment, structural changes occur after repeated ion beam scanning, including defect production\cite{huang2018si}, amorphization\cite{greenzweig2016current}, and ion mixing of materials\cite{rubanov2004fib}, leading to device failure or degradation. Therefore, studies of damage effects are a necessary part of assessing the suitability of novel FIB instruments for applications. \par

Studies on ion damage using commercially available FIB instruments have been widely reported. The Ga liquid metal ion source is the most popular and widely-used ion source due to its good imaging resolution and high material removal rates. Previous study shows that Ga$^+$ irradiation at 30 keV beam energy under normal incidence can lead to an amorphous layer with a thickness exceeding 50 nm with a Ga staining level of 20 at\%  \cite{xiao2013study}. The formation of a Ga concentration layer near  the sample surface is also reported if the ion dose is higher than $1.5\times 10^{17}$ ions/cm$^2$. For light ions, the helium gas field ionization ion source is commonly used \cite{morgan2006introduction}, e.g., to generate nanostructures in nano-film\cite{schmidt2018structurally} or two-dimensional materials\cite{emmrich2016nanopore}. In bulk materials, He$^+$ irradiation usually causes subsurface implantation and damage, such as nano-bubbles in Si substrates \cite{li2019study}. This mainly results from the low sputter yield\cite{livengood2009subsurface} and high diffusivity\cite{hang2014raman} of He$^+$ in silicon. \par

Cold atom ion sources (CAIS) are a relatively new development that can offer diverse ion species as well as high brightness and low energy spread\cite{mcclelland2016bright}, which can potentially reduce ion damage by allowing operation at low beam energy with good resolution. One novel FIB using a Cs$^+$ CAIS was studied in detail\cite{knuffman2013cold} and is now commercialized\cite{zeroktech}. However, experimental studies of ion-induced damage by cold-atom FIBs are still rare. Using a conventional Cs source it was shown that normal incidence Cs$^+$ irradiation at a beam energy of 14.5 keV on Si leads to a 25 nm thick amorphous layer with staining level of around 10 at\% \cite{drezner2016energetic}. 

Employing Rb$^+$ provides an alternative to the Cs$^+$ CAIS and was also shown to enable low energy spread when operated at beam energies under 10 keV\cite{ten2017}. Previous work also demonstrated that Rb$^+$ ions have a good sputtering ability for various substrates\cite{xu2022investigation} at this beam energy. In this paper, we consider the irradiation damage caused by Rb$^+$ ion milling of  Si substrates in nanomachining applications as a further test of suitability. Here, silicon is chosen to be the substrate due to its broad use in semiconductor devices. To characterize the subsurface damage, irradiated regions were cross-sectioned and thinned to produce lamellae. These lamellae were then analyzed by high-resolution TEM combined with energy dispersive X-ray spectroscopy (EDS) to obtain the top-down structural evolution. This work presents new insight into the Rb$^+$ irradiation damage, especially microstructural changes in the amorphous layer.

\section{Experiment}
The Rb$^+$ irradiation tests were performed in a prototype ultracold Rb FIB instrument\cite{xu2022investigation, ten2017}, at a beam energy of 8.5 keV and at a normal incidence. The $<$100$>$ silicon substrates were obtained commercially from Ted Pella Inc with ultrahigh purity. After irradiation, the silicon substrates were transported to a Dual-beam FIB for the lamella preparation.  
\par
To enable TEM characterization of the irradiated silicon substrate, the irradiated regions were first covered by 300 nm of electron-induced Pt deposition and then by 1 $\mu$m of ion-induced Pt deposition as support and protection during lamella preparation in a FEI Nova Nanolab Dual-beam FIB system. The lamella process follows the normal routine. In the final step of polishing, the ion beam energy is set to 5 keV and finally to 2 keV to reduce the Ga$^+$ staining and sidewall amorphization of the lamella.
\par
The lamella samples were characterized in an FEI Titan Cubed 60-300 operated at 300 kV and a JEOL ARM 200F working at 200 kV. Images were acquired both in high-resolution transmission electron microscopy (HRTEM) mode and in high-angle annular dark field (HAADF) scanning transmission electron microscopy (STEM) mode. EDS was performed using the Super-X G2 (Titan) and Centurio SDD (ARM 200F) detectors in STEM mode. The expected EDS measurement accuracy is approximately 1\% and 2\% for heavy and light elements, respectively. Distance measurements on substrates have an estimated statistical uncertainty of typically 0.3 nm as determined by repeated measurements at different positions.

\section{Results and discussion}
\begin{figure}[ht]
    \centering
    \includegraphics[width = \textwidth]{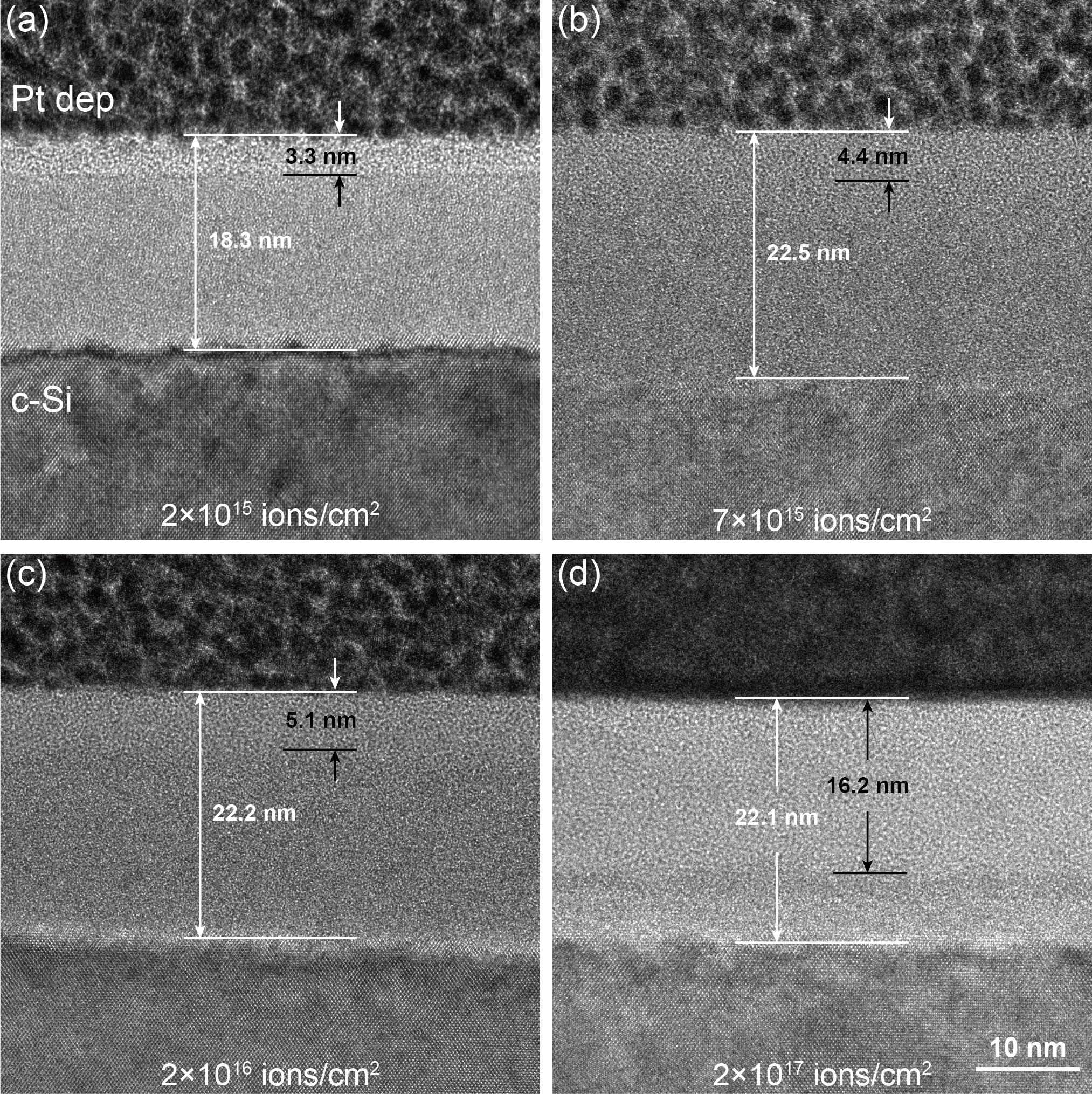}
    \caption{HRTEM images of Si substrates irradiated at normal incidence under various Rb$^+$ ion doses: (a) $2\times10^{15}$ ions/cm$^2$ with 18.3($\pm0.3$) nm damage layer, (b) $7\times10^{15}$ ions/cm$^2$ with 22.5($\pm0.4$) nm damage layer, (c) $2\times10^{16}$ ions/cm$^2$ with 22.2($\pm0.2$) nm damage layer, (d) $2\times10^{17}$ ions/cm$^2$ with 22.1($\pm0.3$) nm damage layer.}
    \label{fig:TEM}
\end{figure}

The Rb$^+$ ion damage study starts with a series of irradiations on Si (100) substrates for various ion doses. Figure\,\ref{fig:TEM} shows the HRTEM images of the irradiated Si samples under four different ion doses. At the lowest dose of  $2\times 10^{15}$ ions/cm$^2$, an amorphous layer of 18.3($\pm0.3$) nm is visible. When the dose is increased to $7\times 10^{15}$ ions/cm$^2$, the thickness is increased to 22.5($\pm0.4$) nm. For even higher ion dose irradiations of $2\times 10^{16}$ and $2\times 10^{17}$ ions/cm$^2$, the damage layer is measured as 22.2($\pm0.2$) and 22.1($\pm0.3$) nm, as shown in Fig.\,\ref{fig:TEM}(c) and (d) respectively. These results show a constant amorphization depth of the silicon substrate by Rb$^+$ ions as a function of dose for all but the lowest dose. At the lowest dose, the damage layer is shown to be around 18.3 nm, as shown in Fig.\,\ref{fig:TEM}(a), which is different from the normal amorphization process that shows a generally increasing thickness in the amorphous layer as described in Drezner \emph{et al} work\cite{drezner2016energetic}. Apart from that, another observation is the stratification in the damage layer. The topmost part of each damaged layer shows increased brightness in the HRTEM images. The thickness of the top layer starts from 3.3($\pm0.1$) nm at the dose of $2\times 10^{15}$ ions/cm$^2$, generally increases to 4.4($\pm0.1$) nm and 5.1($\pm0.1$) nm, and reaches 16.2($\pm0.2$) nm at the highest dose of $2\times 10^{17}$ ions/cm$^2$. It is also noticed that the brighter layer is visible in Fig.\,\ref{fig:TEM}(a) and (d), but not clear to be seen in (b) and (c). Corresponding TEM images at lower magnification are shown additionally in the Appendix in Fig.\,\ref{fig:low mag}, where the amorphous layer with the brighter layer at the top is more apparent.
\par
\begin{figure*}
    \centering
    \includegraphics{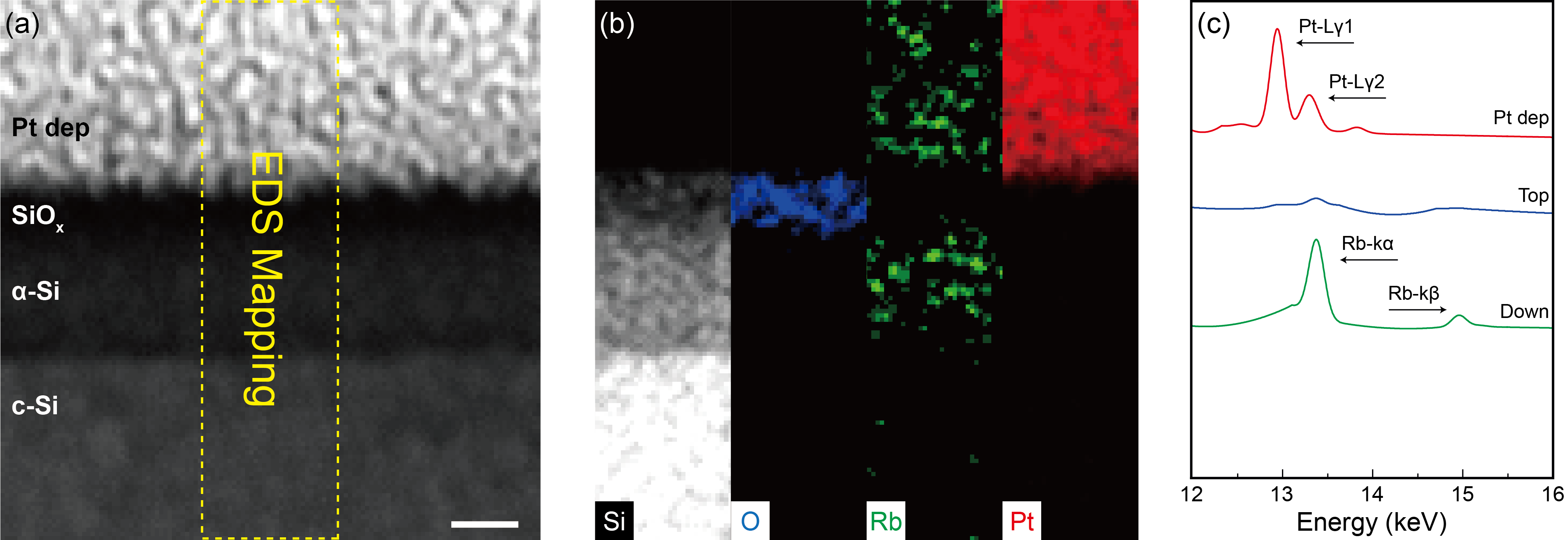}
    \caption{(a) HAADF-STEM image of a cross-section Si sample irradiated with a 2$\times$10$^{16}$ ions/cm$^2$ ion dose. The yellow box inset shows the EDS mapping area. (b) EDS mapping results of the selected area in (a) for 4 elements, white representing Si, blue for O, green for Rb, and red for Pt. (c) The energy spectra for the four layers shown in (a) with the range in 12-16 keV. Scale bar, 10 nm.}
    \label{fig:EDS line}
\end{figure*}

To identify the nature of the two sublayers in the multilayer structure, EDS elemental mappings were acquired in the same areas shown in Fig.\,\ref{fig:TEM}. Here the EDS result of the irradiated region with the ion dose of 2$\times$10$^{16}$ ions/cm$^2$ is shown in Fig.\,\ref{fig:EDS line} as an example. In HAADF-STEM mode, the image brightness scales approximately with the atomic number squared. It can be seen that now the top layer is darker compared to the lower layer as shown in Fig.\,\ref{fig:EDS line}(a), which indicates a small difference in elemental composition between the two layers. Figure\,\ref{fig:EDS line}(b) shows the EDS mapping of 4 elements in the area shown in Fig.\,\ref{fig:EDS line}(a). Silicon can be detected in the top and lower layers, while oxygen is found in the top layer only. It should be noted that rubidium is observed in the lower layer and also seems to be present in the Pt deposition layer, but the latter can be explained by partly overlapping peaks in the EDS spectrum, see Fig.\,\ref{fig:EDS line}(c). For the Rb mapping, the K-lines have been used, as the Rb-L peak (1.69 keV) - although higher in intensity than the Rb K-lines - has a strong overlap with the high-intensity Si K$\alpha$ peak (1.74 keV) from the substrate, making an accurate determination of the Rb content impossible. As an alternative, the Rb K peaks can be used. Fig.\,\ref{fig:EDS line}(c) displays EDS spectra in the energy range of 12-16 keV, corresponding with the four stacked layers shown in Fig.\,\ref{fig:EDS line}(a). Two obvious peaks can be seen in the green curve (corresponding to the lower layer), which represent the Rb-K$\alpha$ (13.37 keV) and K$\beta$ peaks (14.96 keV), respectively. Concerning the red curve (Pt deposition layer), there are also two peaks near 13 keV, identified as Pt-L$\gamma$ peaks. There cannot be any Rb in the Pt dep layer, since the Pt was deposited in the Dual-beam FIB system after the Rb irradiation step. Thus, the overlap between the Pt-L$\gamma$ peaks and Rb-K$\alpha$ peak indeed leads to a wrong assignment of the rubidium distribution in the Pt layer. From the EDS mapping results and spectrum analysis, the top layer of the sample itself only contains silicon and oxygen without any rubidium, while silicon and rubidium are detected in the lower layer. 
\par

\begin{figure}[ht]
    \centering
    \includegraphics{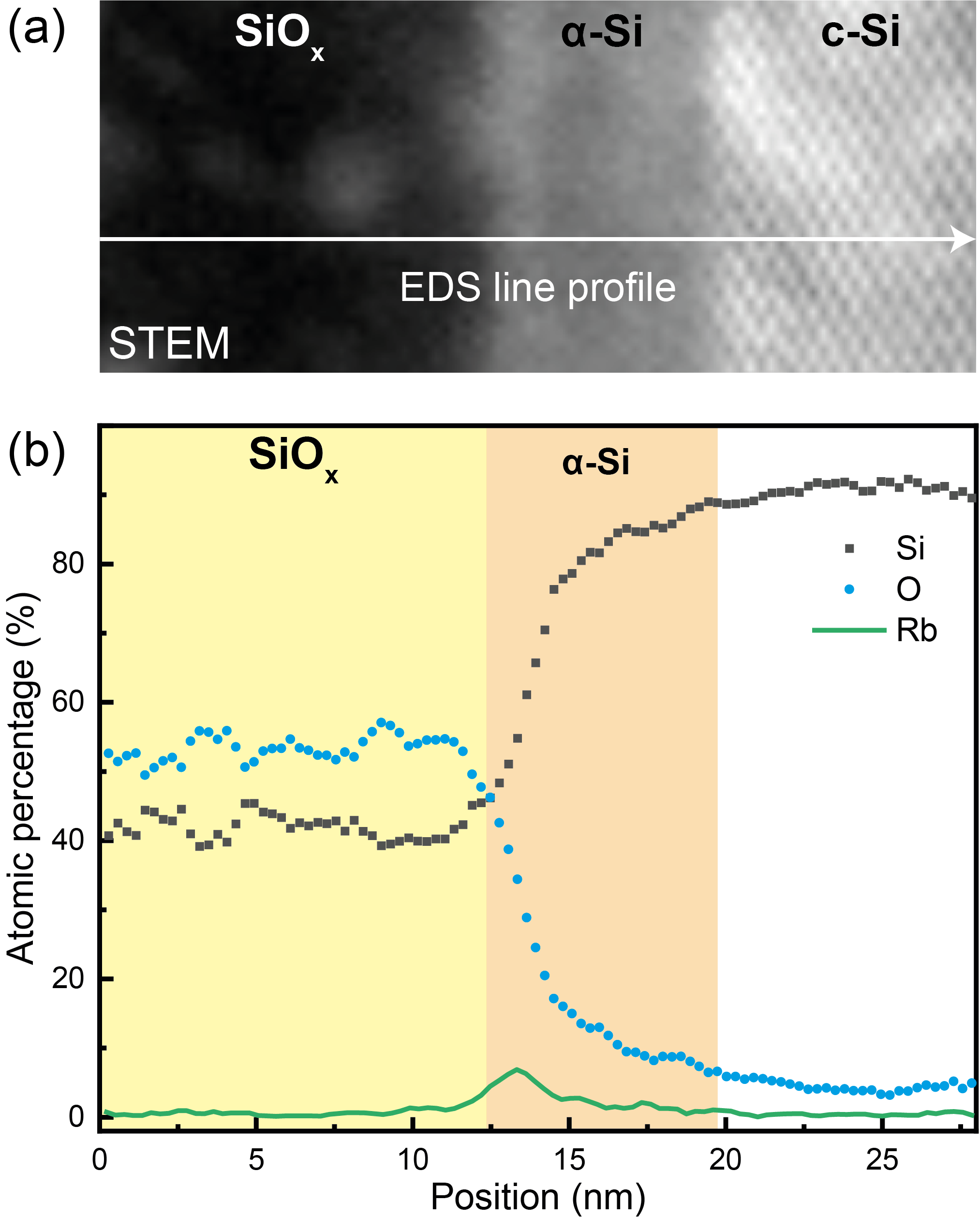}
    \caption{(a) HRTEM image of the top part of an irradiation Si sample at a dose of 3$\times$ 10$^{17}$ ions/cm$^2$; (b) corresponding STEM-HAADF image of (a); (c) Quantitative elemental composition profile across the stack. Black  symbols represents Si, blue is for O, and the green curve is for Rb.}
    \label{fig:EDS no Pt}
\end{figure}

To avoid such erroneous quantification of Rb and Pt, the testing area is chosen to include part of the top layer, lower layer, and crystal silicon (c-Si) substrate without Pt capping, as shown in Fig.\,\ref{fig:EDS no Pt}(a). A quantitative elemental profile, extracted from the EDS mapping of the region shown in Fig.\,\ref{fig:EDS no Pt}(a) is presented in Fig.\,\ref{fig:EDS no Pt}(b). It is observed that in the topmost layer (the left region in Fig.\,\ref{fig:EDS no Pt}(a)), there is almost no Rb, but about 55\% O with around 40\% Si. It is more like an oxidation layer after damage and can be called the "SiO$_x$" layer. The middle region in Fig.\,\ref{fig:EDS no Pt}(a), it has around 90\% Si and low content of O. Thus this layer can be seen as the $\alpha$-Si layer. A Rb peak is shown near the interface between the SiO$_x$ layer and $\alpha$-Si layer with a peak value of 7\% in atomic percentage. Beyond that, an undamaged crystalline Si layer is apparent.\par

\begin{figure}
    \centering
    \includegraphics[width=0.7\textwidth]{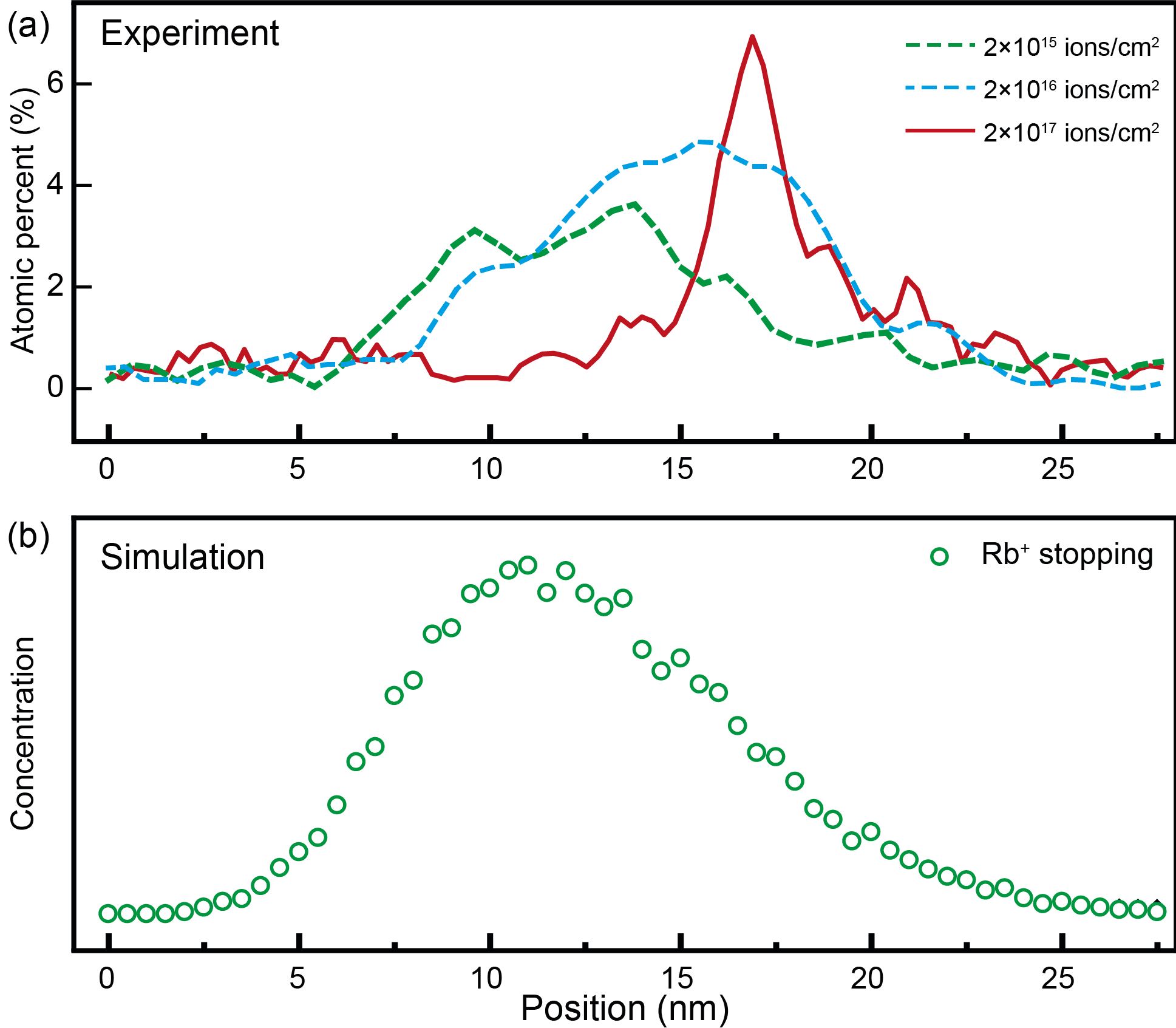}
    \caption{(a) Experimental result of Rb$^+$ distribution at the dose of $2\times10^{15}$ ions/cm$^2$ (green dashed line), $2\times10^{16}$ ions/cm$^2$(blue dashed line), and $2\times10^{17}$ ions/cm$^2$ (red line). (b) SRIM simulation results of the distributions of Rb$^+$ ion stopping sites.}
    \label{fig:SRIM}
\end{figure}

Ion staining is another key factor to characterize ion irradiation damage. Here the Rb concentration profiles for a series of ion doses are presented in Fig.\,\ref{fig:SRIM}(a), as well as the SRIM simulation results (shown in Fig.\,\ref{fig:SRIM}(b)). The zero point position in the figures corresponds to the Pt/Si interface. In general, the Rb peak shifts from the near-surface region to deeper sites as the ion dose increases. At the lowest ion dose of 2$\times$10$^{15}$ ions/cm$^2$ (green dashed curve), the Rb distribution is mainly concentrated in the range of 7-18 nm from the interface (25\%-75\% of peak value). The curve is similar to the SRIM simulation results, where the predicted Rb stopping range is 6-20 nm. For a higher ion dose irradiation of 2$\times$10$^{16}$ ions/cm$^2$, the Rb curve keeps a similar same shape but the peak shifts to around 15 nm from the interface. At the highest ion dose irradiation of 2$\times$10$^{17}$ ions/cm$^2$, the distribution has a sharper peak curve compared to the low ion dose irradiation. The peak shifts from 15 nm to about 17 nm from the interface, and the peak value of Rb content is about 7\%, which can be seen as the stationary Rb$^+$ staining level in Si. The staining level is much lower compared to that of Ga in Si irradiated by Ga$^+$ ions as described in Introduction. Similar work was reported by Drezner \emph{et al}\cite{drezner2016energetic} in which the Cs staining level is measured to be 11.5\% in Si. They suggest the possible explanation for the change in the shape of the distribution could be a high self-sputtering rate of implanted ions by new incident ions.  \par

\begin{figure}
    \centering
    \includegraphics{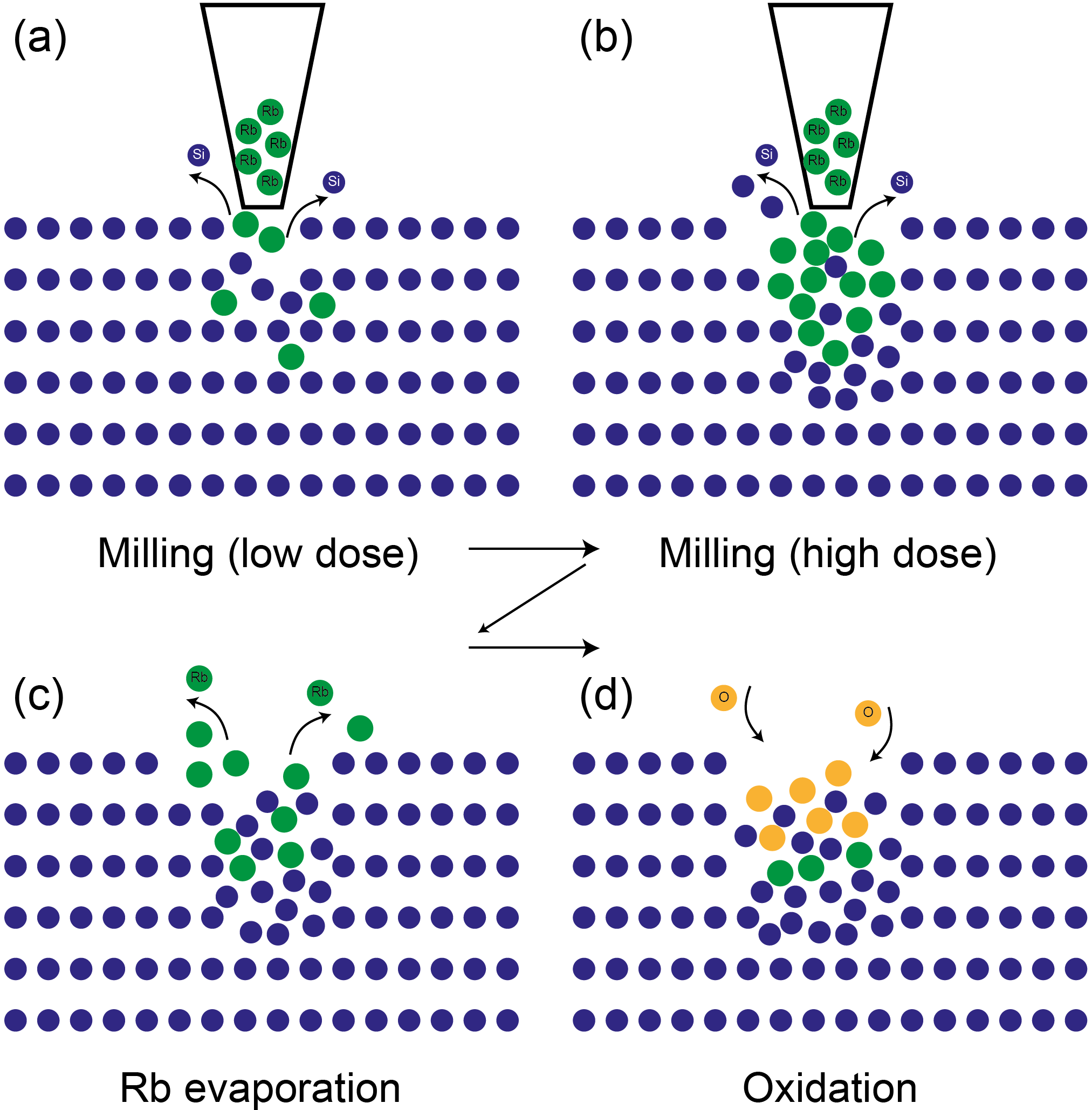}
    \caption{Schematic progress of Rb$^+$ ion damage formation on the Si substrate. Blue represents Si, green is for Rb and orange represents O.}
    \label{fig:schematic}
\end{figure}

In schematic form, Figure\,\ref{fig:schematic} poses a sequence of steps resulting from Rb$^+$ irradiation of a Si substrate to explain the observations. At the start of the ion irradiation, the Rb$^+$ ions penetrate the Si substrate, collide with silicon atoms, and generate damage leading to amorphization. Fig.\,\ref{fig:schematic}(a) shows that at low ion dose, Rb$^+$ ions sputter Si atoms, concentrate in the incident volume, and part of them can reach a deep site from the surface. As predicted by SRIM, the ion range of Rb$^+$ is 12.9 nm, but a small fraction of the Rb$^+$ ions will reach 20 nm depth from the surface in Si. With the increase of impinging Rb$^+$ ions, as shown in Fig.\,\ref{fig:schematic}(b), there is more Rb concentrated in the near-surface region of the Si substrate, part of which is then sputtered and thus removed by the incident ions, while another part is propelled from the near-surface area to a deeper site. Due to the volatility of Rb, it evaporates from the surface after the irradiation (Fig.\,\ref{fig:schematic}(c)). It is known that Rb has a high diffusion in crystalline Si substrate\cite{mccaldin1964alkali}. In addition, the Rb trapped near the surface can also get enough thermal energy to overcome a surface barrier potential during repeated scanning of the irradiating ion beam\cite{atutov2017}. A high dose furthermore leads to a more porous structure of the Si substrate, which enhances Rb evaporation in deeper regions. This leads to the shift of Rb concentration observed in Fig.\,\ref{fig:SRIM}. The formation of the SiO$_x$ layer is then mainly caused by the introduction of oxygen to the irradiated area as shown in Fig.\,\ref{fig:SRIM}(d). One possibility is the presence of residual oxygen gas in the FIB vacuum chamber during milling, but more likely is that it results from air exposure during the sample transportation from the Rb FIB to the Dualbeam FIB. Such an oxidation process might occur preferably in the porous Si regions and form a barrier layer to prevent outer oxygen penetration and rubidium evaporation.  The observations would indicate that nanomachining with Rb$^+$ ions of a Si-based substrate requires a method to avoid the oxidation process if oxidation is undesirable.

\section{Conclusions}
A series of Rb$^+$ irradiation experiments at 8.5 keV beam energy under normal incidence has shown a 22.2($\pm$0.3) nm thick damage layer in silicon substrates. A multi-layer damage structure is observed consisting of an oxidation layer and an amorphous silicon layer. With increasing of ion dose, the thickness of the oxidation layer also increases, from 3.3($\pm0.1$) nm ($3\times10^{17}$ ions/cm$^2$) to 4.4($\pm0.1$) nm and 5.1($\pm0.1$) nm, and finally to at 16.2($\pm0.2$) nm ($3\times10^{17}$ ions/cm$^2$). The measured Rb$^+$ staining level in Si is 7($\pm$1)\% in atomic weight. This damage layer thickness and ion staining level is comparable to what was observed for Cs$^+$ irradiation \cite{drezner2016energetic} under somewhat similar circumstances. Both layer thickness and staining level are reduced when compared to damage and staining resulting from 30 keV Ga$^+$ beams. In general, therefore, Rb$^+$ and Cs$^+$ CAIS operated at low beam energy can be a solution to decrease ion damage on Si. Formation of an oxidation layer can however be expected.

\begin{acknowledgments}
This work is part of the project Next-Generation Focused Ion Beam (No.16178) of the research programme Applied and Engineering Sciences (TTW) which is (partly) financed by the Dutch Research Council (NWO). The authors are (partly) members of the FIT4NANO COST Action CA19140. Solliance and the Dutch province of Noord-Brabant are acknowledged for funding the STEM facility. The authors acknowledge helpful discussions with Greg Schwind, Chad Rue, Yuval Greenzweig, and Peter Graat.
\end{acknowledgments}

\section*{Author Declarations}
\subsection*{Conflict of Interest}
The authors have no conflicts to disclose.

\section*{Data Availability Statement}

The data that support the findings of this study are available from the corresponding author upon reasonable request.

\clearpage
\appendix

\section{Rb ion damage at low magnification}

\begin{figure}[ht]
    \centering
    \includegraphics{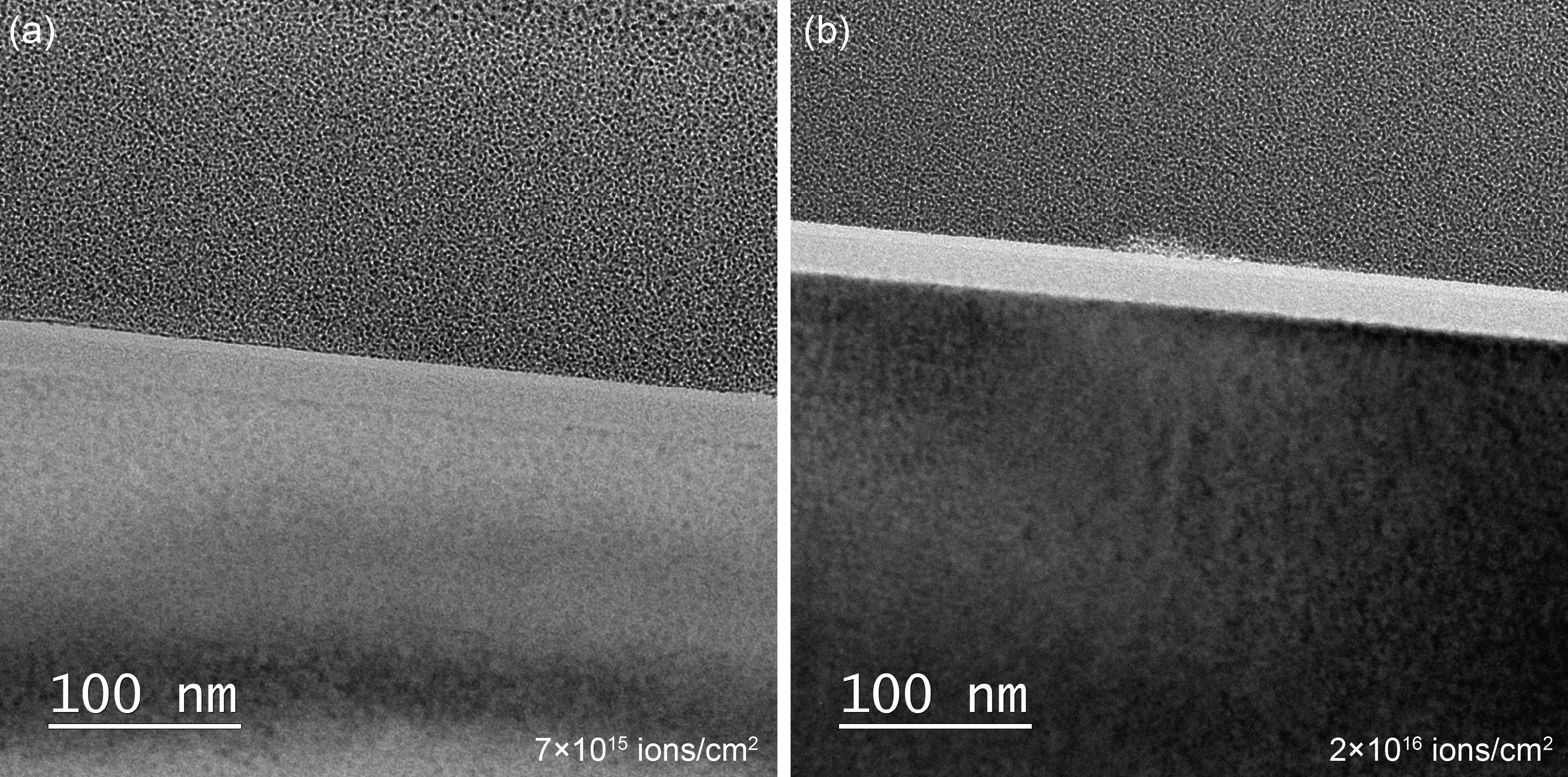}
    \caption{Low magnification TEM images of Si substrates irradiated at normal incidence under Rb$^+$ ion doses: (a) $7\times10^{15}$ ions/cm$^2$ and (b) $2\times10^{16}$ ions/cm$^2$}
    \label{fig:low mag}
\end{figure}

\clearpage
\nocite{*}
\bibliography{ref}

\end{document}